\documentclass[aps,pre,showpacs,twocolumn]{revtex4}
\usepackage{graphicx}  
\usepackage{dcolumn} 
\usepackage{bm} 

\begin{document}

\title{Theory of the Maxwell Pressure Tensor and the Tension in a Water Bridge}

\author{A. Widom}
\author{J. Swain}
\author{J. Silverberg}
\affiliation{Physics Department, Northeastern University, Boston MA 02115}
\author{S. Sivasubramanian}
\affiliation{NSF Center for High-rate Nanomanufacturing, Northeastern University, Boston MA 02115}
\author{Y.N. Srivastava}
\affiliation{Physics Department \& INFN, University of Perugia, Perugia IT}

\begin{abstract}

A water bridge refers to an experimental ``flexible cable'' made up of pure deionized water 
which can hang across two supports maintained with a sufficiently large voltage difference.
The resulting electric fields within the deionized water flexible cable, maintain a tension
which sustains the water against the downward force of gravity. A detailed calculation of 
the water bridge tension will be provided in terms of the Maxwell pressure tensor in a 
dielectric fluid medium. General properties of the dielectric liquid pressure tensor are 
discussed along with unusual features of dielectric fluid Bernoulli flows in an electric field.
Analogies between dielectric fluid Bernoulli flows in strong electric fields and quantum 
Bernoulli flows in superfluids are explored.

\end{abstract}


\maketitle

\section{Introduction \label{intro}}

Recent observations\cite{Fuchs:2007,Fuchs:2008,Hand:2007} have been made of water bridges  
stretched across supports which are maintained at large voltage differences. A water bridge 
is a ``flexible cable'' made up of pure deionized water which has an electric field 
\begin{math} {\bf E}  \end{math} in virtue of an applied voltage across the supports at the 
ends of the fluid cable. Previously to this work, it was not fully understood what forces 
hold up the water bridge against the force of gravity. It will here be shown that the forces 
responsible for holding up the water bridge follow from the Maxwell electric field 
pressure tensor in dielectric polar fluids. In particular, the water bridge viewed as a 
flexible cable has an electric field induced tension sufficiently large so as to explain 
the water bridge support. In discussing the general theory of the pressure tensor for isotropic 
dielectric polar liquids, such as water, hydrostatics and adiabatic hydrodynamics shall also be 
explored.

In Sec.\ref{thermo} the thermodynamic laws applied to a fluid polar dielectric   
are described in detail. It is shown that in the presence of an electric field, 
there are two different thermodynamic pressures, \begin{math} P \end{math} and 
\begin{math} \tilde{P} \end{math}. These turn out to be eigenvalues of the 
full pressure tensor as discussed in Sec.\ref{press}. If an infinitesimal surface area 
\begin{math} \delta A_\perp \end{math} has a normal perpendicular 
to the electric field lines, then the pressure force is 
\begin{math} P\delta A_\perp \end{math}. If an infinitesimal surface area 
\begin{math} \delta A_\parallel \end{math} has a normal parallel 
to the electric field lines, then the pressure force is 
\begin{math} \tilde{P}\delta A_\parallel \end{math}.
It is shown that these results completely characterize the pressure 
tensor. 

In Sec.\ref{cyl}, we compute the tension in a fluid dielectric cylinder. The physics 
of that calculation is easily explained. Consider a simple cylinder of length 
\begin{math} L \end{math} and cross sectional area \begin{math} A=\pi R^2 \end{math}. 
Suppose that a spatially uniform electric field \begin{math} {\bf E} \end{math} exists 
in a direction parallel to the cylinder axis. Since the tangential component of the 
electric field is continuous, the electric field inside the cylinder is the same as 
the electric field outside the cylinder. On the other hand, the displacement field 
\begin{math} {\bf D} =\varepsilon {\bf E} \end{math} is discontinuous at the endpoints 
of the cylinder; i.e. the ends of the cylinder have a charge density of     
\begin{math} \pm \sigma \end{math} wherein  
\begin{equation}
4\pi \sigma =\Delta D=(\varepsilon -1)E=4\pi \frac{Q}{A}\ .
\label{intro1}
\end{equation}
The end of the cylinder which is at the tail of the electric field vector has charge 
\begin{math} +Q \end{math} while the end of the cylinder which is at the arrow of the 
electric field vector has charge \begin{math} -Q \end{math}. The tension in the cylinder 
is then evidently given by \begin{math} \tau=QE \end{math} which in virtue of 
Eq.(\ref{intro1}) reads   
\begin{equation}
\frac{\tau }{A}=\left[\frac{\varepsilon-1}{4\pi }\right] E^2\ .
\label{intro2}
\end{equation}
In what follows, the stress Eq.(\ref{intro2}) will be rigorously derived from the Maxwell 
pressure tensor within the cylinder. The ratio of the tension to cylinder weight, 
\begin{math} Mg=\rho ALg  \end{math}, obeys 
\begin{equation}
\frac{\tau }{Mg}=\left[\frac{(\varepsilon-1)E^2}{4\pi \rho gL}\right], 
\label{intro3}
\end{equation}
wherein \begin{math} \rho \end{math} and \begin{math} L \end{math} represent, respectively, 
the mass density and length of the cylinder. Although we have employed free end boundary 
conditions to the cylinder, the tension \begin{math} \tau \end{math} is a local stress quantity 
independent of global boundary conditions. In Sec.\ref{hfc}, we exhibit a plot of the hanging 
water bridge flexible cable for experimental values of the parameters in Eq.(\ref{intro3}) 
and find the agreement between theory and experiment to be satisfactory.

In Sec.\ref{efh}, the hydrostatics of the dielectric fluid in an electric field is explored. 
The crucial quantity of interest is the mean isothermal polarization per molecule which obeys  
\begin{equation}
{\bf p}=\alpha_T {\bf E}\ \ \ \Rightarrow 
\ \ \ 4\pi \alpha_T =m\left(\frac{\partial \varepsilon }{\partial \rho } \right)_{T,{\bf E}},
\label{intro4}
\end{equation}
wherein \begin{math} m \end{math} is the mass of a single molecule and 
\begin{math} \alpha_T \end{math} is the isothermal polarizability. For the water liquid and 
vapor phases, the molecular polarizabilities, respectively, obey 
\begin{math} \alpha_T^{gas}\ll \alpha_T^{liquid} \end{math}. Fluid dielectric films in strong 
electric fields adsorbed on insulating walls tend to swell to a large thickness. 
In Sec.\ref{eff}, the theory of dielectric polar fluid Bernoulli flows in strong electric 
fields is discussed. Together with the effect of film thickening, it turns out that the film 
is capable of crawling up an insulating container wall against the force of gravity and flow 
as in a siphon over the top of the wall and down the other side. In the concluding 
Sec.\ref{conc}, the role of Bernoulli flows in forming water bridges will be discussed. Analogies 
between dielectric polar liquid Bernoulli flows in strong electric fields and quantum 
Bernoulli flows in superfluids will be explored. These analogies include the ability of films 
to climb walls against gravitational forces and even to pass over the tops of these walls over to 
the other side.

\section{Thermodynamic Arguments \label{thermo}}  

Let \begin{math} \tilde{f}(\rho ,T,{\bf D})  \end{math} represent the Helmholtz free 
energy per unit volume for a dielectric fluid of mass density \begin{math} \rho \end{math}, 
temperature \begin{math} T \end{math} and Maxwell displacement field 
\begin{math} {\bf D} \end{math};
\begin{equation}
d\tilde{f}=-sdT+\zeta d\rho +\frac{1}{4\pi}{\bf E}\cdot d{\bf D},
\label{thermo1}
\end{equation}
wherein \begin{math} s \end{math} is the entropy per unit volume, 
\begin{math} \zeta \end{math} is the chemical potential per unit mass 
and \begin{math} {\bf E} \end{math} is the electric field. 
The thermodynamic pressure which follows from Eq.(\ref{thermo1}),
\begin{eqnarray}
\tilde{P}=\zeta \rho-\tilde{f},
\nonumber \\ 
d\tilde{P}=sdT+\rho d\zeta-\frac{1}{4\pi}{\bf E}\cdot d{\bf D}. 
\label{thermo2}
\end{eqnarray}
On the other hand, one may employ the free energy 
\begin{eqnarray}
f=\tilde{f}-\frac{1}{4\pi }{\bf E}\cdot {\bf D},
\nonumber \\ 
df=-sdT+\zeta d\rho -\frac{1}{4\pi}{\bf D}\cdot d{\bf E}, 
\label{thermo3}
\end{eqnarray}
yielding the pressure 
\begin{eqnarray}
P=\zeta \rho-f,
\nonumber \\ 
dP=sdT+\rho d\zeta+\frac{1}{4\pi}{\bf D}\cdot d{\bf E}. 
\label{thermo4}
\end{eqnarray}
The two different pressures obey 
\begin{equation}
P=\tilde{P}+\frac{1}{4\pi }{\bf E\cdot D}.
\label{thermo5}
\end{equation}
For a dielectric fluid in an electric field, isotropy dictates 
that \begin{math} {\bf D}  \end{math} be parallel to \begin{math} {\bf E}  \end{math} 
even if the detailed equations of state are non-linear, i.e. isotropy yields a free energy 
of the form 
\begin{equation}
f(\rho ,T, {\bf E})\equiv f(\rho ,T, E^2)
\label{thermo6}
\end{equation}
so that Eqs.(\ref{thermo3}) and (\ref{thermo6}) imply  
\begin{equation}
{\bf D}=\varepsilon {\bf E},
\label{thermo7}
\end{equation}
wherein 
\begin{math} \varepsilon (\rho ,T, E^2)=
-8\pi \partial f(\rho ,T, E^2)/\partial (E^2) \end{math}.
Thus, the two possible fluid pressures in Eq.(\ref{thermo5}) obey 
\begin{equation}
\tilde{P}=P-\frac{\varepsilon}{4\pi }E^2.
\label{thermo8}
\end{equation}
It may at first glance appear strange that there are two physically different thermodynamic 
pressures in a dielectric fluid subject to an electric field. However, the situation may be 
clarified when it is realized that due to the electric field, the pressure is in reality a 
{\em tensor}. 

\section{Pressure Tensor \label{press}}  

In order to compute the Maxwell pressure tensor in a fluid dielectric, imagine that the fluid 
undergoes a strain wherein a fluid particle at point \begin{math} {\bf r} \end{math} 
is sent to the new point \begin{math} {\bf r}^\prime  \end{math}. Such a transformation induces 
a strained length scale \begin{math} ds^2=d{\bf r}^\prime \cdot d{\bf r}^\prime \end{math} 
described by a metric tensor  
\begin{equation}
ds^2=g_{ij}({\bf r})dr^idr^j.
\label{press1}
\end{equation}
The pressure tensor \begin{math} P_{ij} \end{math} is then described in terms of the free energy 
change due to a metric strain 
\begin{equation}
\delta F=\frac{1}{2}\int P_{ij}\delta g^{ij}dV.
\label{press2}
\end{equation}
The free energy variation is described by
\begin{equation}
\delta F=\delta \int fdV=\int \delta f dV+\int f\delta dV.
\label{press3}
\end{equation}
In detail, the volume element of the strained fluid is determined by 
\begin{math} g({\bf r})=\det [g_{ij}({\bf r})]  \end{math} via 
\begin{equation}
dV=\sqrt{g({\bf r})}d^3{\bf r}\ \ \ \Rightarrow 
\ \ \ \delta dV=-\frac{1}{2}g_{ij}\delta g^{ij}dV.
\label{press4}
\end{equation}
The volume variational Eq.(\ref{press4}) together with mass conservation 
in turn implies a mass density variation 
\begin{equation}
\delta \rho =\frac{1}{2}g_{ij}\delta g^{ij}\rho .
\label{press5}
\end{equation}
Furthermore, the change in the magnitude of the electric field is 
\begin{equation}
\delta E^2 =\delta g^{ij}E_iE_j .
\label{press6}
\end{equation}
We then employ  
\begin{eqnarray}
\delta f=\left(\frac{\partial f}{\partial \rho }\right)_{T,E^2}\delta \rho 
+\left(\frac{\partial f}{\partial E^2 }\right)_{T,\rho}\delta E^2 ,
\nonumber \\ 
\delta f=\zeta \delta \rho -\frac{\epsilon}{8\pi }\delta E^2.
\label{press7}
\end{eqnarray}
In virtue of Eqs.(\ref{press3})-(\ref{press7}), 
\begin{equation}
\delta F=\frac{1}{2}\int \delta g^{ij}
\left[g_{ij}(\zeta \rho -f)-\frac{\varepsilon }{4\pi }E_iE_j\right]dV.
\label{press8}
\end{equation}
From Eqs(\ref{thermo4}), (\ref{press2}) and (\ref{press8}) we have the final 
form\cite{Becker:1964,Landau:1960} for the pressure tensor 
\begin{equation}
P_{ij}=Pg_{ij}-\frac{\varepsilon }{4\pi }E_iE_j.
\label{press9}
\end{equation}
In dyadic notation, the pressure tensor is given by 
\begin{eqnarray}
{\sf P}=P{\sf 1}-\frac{\varepsilon }{4\pi }{\sf EE},
\nonumber \\ 
{\sf P}=P{\sf 1}-\frac{\varepsilon E^2}{4\pi }{\sf nn}
\ \ {\rm with}\ \ {\sf n}=\frac{\sf E}{E},
\nonumber \\ 
{\sf P}=P{\sf 1}+(\tilde{P}-P){\sf nn},
\label{press10}
\end{eqnarray}
wherein Eq.(\ref{thermo8}) has been invoked.

It is now clear as to why there are two thermodynamic pressures, 
\begin{math} P \end{math} and \begin{math} \tilde{P} \end{math}, in 
Sec.\ref{thermo}. For an infinitesimal surface area 
\begin{math} \delta A_\perp \end{math} whose normal is perpendicular 
to the electric field lines, the pressure force is 
\begin{math} P\delta A_\perp \end{math}. For an infinitesimal surface area 
\begin{math} \delta A_\parallel \end{math} whose normal is parallel 
to the electric field lines, the pressure force is 
\begin{math} \tilde{P}\delta A_\parallel \end{math}.
These results describe completely when to use the pressure 
\begin{math} P \end{math} and when to use the pressure 
\begin{math} \tilde{P} \end{math}.

\section{Tension in a Fluid Cylinder \label{cyl}}

For a fluid dielectric cylinder of length \begin{math} L \end{math} and cross 
sectional area \begin{math} A \end{math} in a uniform electric field parallel to the axis, 
let us consider the work done in changing the volume \begin{math} V=LA \end{math} via  
\begin{equation}
dV=AdL+LdA.
\label{cyl1}
\end{equation}
Employing the pressure tensor Eq.(\ref{press10}), one finds that both pressures, 
\begin{math} \tilde{P}  \end{math} and \begin{math} P  \end{math} are required 
\begin{equation}
dW = -\tilde{P}AdL-PLdA.
\label{cyl2}
\end{equation}
When a fluid is stretched at constant volume \begin{math} dV=AdL+LdA=0 \end{math} 
\begin{equation}
-LdA=AdL\ \ \ \Rightarrow \ \ \ dW =(P-\tilde{P})AdL=\tilde{\tau }dL.
\label{cyl3}
\end{equation}
The effective tension is thereby 
\begin{equation}
\tilde{\tau}=(P-\tilde{P})A=\frac{\varepsilon E^2A}{4\pi },
\label{cyl4}
\end{equation}
wherein Eq.(\ref{thermo8}) has been invoked. Subtracting the tension that would be 
present for an electric field in the vacuum, 
\begin{math} \tau =\tilde{\tau }-\tau_{vac}  \end{math}, 
yields our final result for the tension,
\begin{equation}
\tau =\frac{(\varepsilon -1)E^2A}{4\pi }
\label{cyl5}
\end{equation}
in agreement with Eq.(\ref{intro2}) of Sec.\ref{intro}.

\section{Hanging Flexible Cable  \label{hfc}}

\begin{figure}[tp]
\scalebox {0.3}{\includegraphics{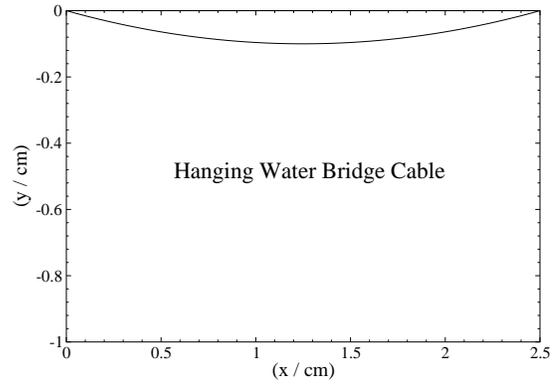}}
\caption{Employing the experimental example\cite{Fuchs:2007,Fuchs:2008} 
wherein the supports are separated by $2.5\ {\rm cm}$ and with $\rho =1\ {\rm gm/cm^3}$, 
$g=980\ {\rm cm/sec^2}$, $\varepsilon =80$ and $E=33\ {\rm Gauss}$, we plot the hanging 
water bridge flexible cable. The maximum hanging dip at the center of the water bridge 
is $h=0.1\ {\rm cm}$. The agreement between experiment and the theoretical 
Eqs.(\ref{intro3}) and (\ref{hfc1}) is satisfactory.}
\label{Fig1}
\end{figure}

The water bridge consists of a flexible fluid cable which can be suspended by its 
endpoints. There is a slight sag in the cable as befits the equilibrium of the total 
gravitational force \begin{math} Mg \end{math} downward and the total Maxwell tension 
force \begin{math} 2\tau \sin \theta  \end{math} upward wherein \begin{math} \theta  \end{math} 
is the angle between the cable tangent at the support and the horizontal;     
\begin{equation}
Mg=2\tau \sin\theta \ \ \ \Rightarrow \ \ \ \sin \theta =\frac{2\pi \rho g L}{(\varepsilon -1)E^2}
\label{hfc1}
\end{equation}
wherein Eq.(\ref{intro3}) has been invoked. Eq.(\ref{hfc1}) allows for the theoretical computation 
of \begin{math} \theta \end{math} in terms of experimental dielectric constants, mass densities, 
water bridge lengths and electric fields. With regard to the electric field, one notes the {\em exact} 
conversion between the Gaussian units here employed and the usual engineering units, 
\begin{equation}
1\ {\rm Gauss}\equiv 299.792458\ {\rm volt/cm}.  
\label{hfc2}
\end{equation}
The agreement between theory and experiment in predicting the slight hang of the water bridge as 
a flexible cable, as in the above Fig.\ref{Fig1}, is satisfactory.

\section{Hydrostatics in an Electric field \label{efh}}

The force density on the fluid as described by the pressure tensor Eq.(\ref{press10}) is given by
\begin{equation}
{\bf f}=-div{\sf P}=-{\bf grad}P+div\left(\frac{\varepsilon {\sf EE}}{4\pi}\right).
\label{efh1}
\end{equation}
Within the bulk liquid \begin{math} div{\bf D}=div(\varepsilon {\bf E})=0 \end{math} so that
Eq.(\ref{efh1}) thereby reads
\begin{equation}
{\bf f}=-{\bf grad}P+\frac{\varepsilon}{4\pi}({\bf E\cdot grad}){\bf E}.
\label{efh2}
\end{equation}
On the other hand, from the Gibbs-Duhem Eq.(\ref{thermo4}) under equilibrium isothermal 
conditions \begin{math} dT=0  \end{math},
\begin{equation}
{\bf grad}P=\rho {\bf grad}\zeta +\frac{\varepsilon}{4\pi}({\bf E\cdot grad}){\bf E}
\label{efh3}
\end{equation}
so that the force per unit volume can be computed from the chemical potential per 
unit mass; i.e.
\begin{equation}
{\bf f}=\rho {\bf grad}\zeta .
\label{efh4}
\end{equation}
Under a Newtonian gravitational field  
\begin{equation}
{\bf g}=-{\bf grad}\Phi ,
\label{efh5}
\end{equation}
one finds the total force density equilibrium condition
\begin{equation}
{\bf f}+\rho {\bf g}=-{\bf grad}(\zeta  + \Phi)=0
\label{efh6}
\end{equation}
yielding the uniform chemical potential condition 
\begin{equation}
\mu=m[\zeta (\rho ,T,E^2) + \Phi ]={\rm const.}
\label{efh7}
\end{equation}
To compute the electric field dependence of the chemical potential per unit mass,  
one may apply a Maxwell relation to the thermodynamic Eq.(\ref{thermo3}) which 
reads 
\begin{equation}
\left(\frac{\partial \zeta }{\partial {\bf E}}\right)_{T,\rho }=
-\frac{1}{4\pi }\left(\frac{\partial {\bf D}}{\partial \rho}\right)_{T,{\bf E}}
=-\frac{\alpha_T}{m}{\bf E},
\label{efh8}
\end{equation}
wherein the mean molecular dipole moment \begin{math} {\bf p}=\alpha_T {\bf E} \end{math} 
defines the polarizability \begin{math} \alpha_T \end{math} as in Eq.(\ref{intro4}).
Integrating Eq.(\ref{efh8}) completes the calculation of the chemical potential
\begin{eqnarray}
\zeta (\rho ,T, {\bf E})=\zeta_0 (\rho ,T)- 
\frac{1}{2m}\int_0^{E^2}\alpha_T((\rho ,T,F^2))d(F^2),
\nonumber \\ 
\zeta (\rho ,T, {\bf E})=\zeta_0(\rho ,T)-\frac{\alpha_{T}(\rho ,T,E^2=0)}{2m}E^2
+\ldots \ .   
\label{efh9}
\end{eqnarray}
The central Eq.(\ref{efh9}) of this section implies that the chemical potential is 
lowered when strong electric fields are applied.

It follows from Eq.(\ref{efh9}) that the application of an electric field lowers the 
chemical potential of films of water adsorbed on insulating substrates such as  
glass which is often employed in physical chemistry experiments. When the chemical 
potential of a liquid film is lowered, the water film thickness increases; e.g.  
in the presence of electric fields, water in a glass beaker will have a film which 
appears to climb higher up the walls than would be possible in the zero electric 
field case. The fabrication of a water bridge begins by applying a potential 
difference across the water contained in two neighboring glass beakers which just touch 
each other. One expects and finds experimentally a thickening film layer all around 
the points at which the water horizontal surfaces meet the two beaker walls. The water 
climbs the walls of both beakers and near the touch point of the beakers splashes 
over the tops. When the hydrostatic calm after the splash begins, a bridge is formed 
at the point wherein the beakers just touch. A longer water bridge is formed after slowly 
separating the beakers. Let us now turn to the hydrodynamic features of the polar liquid 
flows in a strong electric field.

\section{Bernoulli Flows in Strong Fields \label{eff}}  

We here employ the usual notion of a fluid derivative operator, 
\begin{equation}
\frac{d}{dt}=\frac{\partial}{\partial t}+({\bf v\cdot grad}),
\label{eff1}
\end{equation}
which expresses the time rate of change operator as seen by an observer 
moving locally with the fluid velocity \begin{math} {\bf v} \end{math}. 
For example, mass conservation reads\cite{Landau:1987} 
\begin{equation}
\frac{d\rho }{dt}=-\rho \ div{\bf v}.
\label{eff2}
\end{equation}
Bernoulli flows are adiabatic, i.e. viscous entropy production is ignored. 
Conservation of energy then amounts to a local {\em entropy} conservation 
law\cite{Landau:1987}    
\begin{equation}
\frac{ds}{dt}=-s\ div{\bf v}.
\label{eff3}
\end{equation}
wherein \begin{math} s \end{math} is the entropy per unit volume. 
From Eqs.(\ref{eff1}) and (\ref{eff2}) it follows that 
\begin{equation}
\frac{d}{dt}\left(\frac{s}{\rho }\right)=
\frac{1}{\rho^2 }\left(\rho \frac{ds }{dt}-s\frac{d\rho  }{dt}\right)=0,
\label{eff4}
\end{equation}
which implies a conservation law for the entropy per unit mass 
\begin{equation}
s^*\equiv \frac{s}{\rho }\ \ \ \Rightarrow 
\ \ \ \frac{ds^*}{dt}=0.
\label{eff5}
\end{equation}
It is here that the enthalpy per unit mass \begin{math} w \end{math} makes it's 
way into the adiabatic polar dielectric liquid Bernoulli flows in an electric field; 
\begin{eqnarray}
\zeta =w-Ts^*, 
\nonumber \\ 
dw=Tds^*+\frac{1}{\rho }dP-\frac{1}{4\pi \rho }{\bf D}\cdot d{\bf E},
\label{eff6}
\end{eqnarray}
wherein Eq.(\ref{thermo4}) has been invoked.
For an adiabatic flow with \begin{math} ds^*=0 \end{math} as in Eq.(\ref{eff5}), 
Eq.(\ref{eff6}) implies 
\begin{equation}
\rho {\bf grad}w={\bf grad}P-\frac{1}{4\pi }({\bf D\cdot grad}){\bf E}.
\label{eff7}
\end{equation}
Thus, the Maxwell pressure tensor force per unit volume Eq.(\ref{efh2}) in an 
adiabatic Bernoulli flow reads 
\begin{equation}
{\bf f}=-\rho \ {\bf grad}w.
\label{eff8}
\end{equation} 
The dynamical equation of motions which accounts for momentum conservation 
then becomes 
\begin{eqnarray}
\rho \frac{d{\bf v}}{dt}={\bf f}+\rho {\bf g},
\nonumber \\ 
\frac{d{\bf v}}{dt}=-{\bf grad}(w+\Phi ),
\label{eff9}
\end{eqnarray} 
wherein the gravitational field Eq.(\ref{efh5}) has been taken into account.
Vorticity 
\begin{equation}
{\bf \Omega }=curl {\bf v}
\label{eff10}
\end{equation}
makes an appearance in virtue of the acceleration identities 
\begin{eqnarray}
\frac{d{\bf v}}{dt}=\frac{\partial {\bf v}}{\partial t}
+({\bf v \cdot grad}){\bf v},
\nonumber \\ 
\frac{d{\bf v}}{dt}=\frac{\partial {\bf v}}{\partial t}+
{\bf \Omega \times v}-\frac{1}{2}{\bf grad}(v^2),
\label{eff11}
\end{eqnarray}
which allow us to write Eq.(\ref{eff9}) as 
\begin{equation}
\frac{\partial {\bf v}}{\partial t}+
{\bf \Omega \times v}=-{\bf grad}\left(w+\Phi +\frac{1}{2}v^2\right).
\label{eff12}
\end{equation}
Employing the curl of Eq.(\ref{eff12}) and using Eq.(\ref{eff10}) implies the equation 
of motion for vorticity. It is 
\begin{eqnarray}
\frac{\partial {\bf \Omega }}{\partial t}+curl({\bf \Omega \times v})=0,
\nonumber \\ 
\frac{d{\bf \Omega }}{dt}=({\bf \Omega \cdot grad}){\bf v}-{\bf \Omega }(div {\bf v}).
\label{eff13}
\end{eqnarray}
If at a given initial time the Bernoulli flow is irrotational, i.e. 
\begin{math} {\bf \Omega}=0 \end{math}, then at all later times in accordance with 
Eq.(\ref{eff13}) the flow will remain irrotational. Eq.(\ref{eff12}) then reads\cite{Landau:1987}  
\begin{eqnarray}
{\bf v}={\bf grad}\varphi , 
\nonumber \\ 
\frac{\partial \varphi}{\partial t}+\frac{1}{2}|{\bf grad}\varphi |^2+w+\Phi =0,
\nonumber \\ 
\frac{\partial {\bf v}}{\partial t} 
=-{\bf grad}\left(w+\Phi +\frac{1}{2}v^2\right).
\label{eff14}
\end{eqnarray}
The complete set of equations for a steady state Bernoulli flow in a strong electrostatic 
field \begin{math} {\bf E}  \end{math} and in a uniform gravitational field 
\begin{math} {\bf g} \end{math} in the negative z-direction then follows from 
Eq.(\ref{eff14}) as 
\begin{eqnarray}
curl{\bf E}=0,
\nonumber \\ 
div{\bf D}=div(\varepsilon {\bf E})=0,
\nonumber \\ 
w(s^*,P,{\bf E})+\frac{1}{2}|{\bf v}|^2+gz={\rm const}.
\label{eff15}
\end{eqnarray}
The only difference between the normal steady state Bernoulli fluid flows 
in Eq.(\ref{eff15}) and the usual case for \begin{math} {\bf E}=0 \end{math} 
resides in the electric field contributions to the enthalpy per unit mass 
\begin{math} w \end{math}. It is here useful to introduce a new thermodynamic 
potential per unit mass \begin{math} \varpi \end{math} obeying 
\begin{eqnarray}
w =\varpi +\frac{P}{\rho}, 
\label{eff16} \\ 
d\varpi =Tds^* +\frac{P}{\rho^2}d\rho -\frac{1}{4\pi \rho }{\bf D}\cdot d{\bf E},
\nonumber \\ 
\varpi (s^*,\rho ,E^2)=\varpi_0 (s^*,\rho )-
\nonumber \\ 
\frac{1}{8\pi \rho } \int_0^{E^2} \varepsilon (s^*,\rho ,F^2)d(F^2),
\nonumber \\ 
\varpi (s^*,\rho ,E^2) = \varpi_0 (s^*,\rho )
-\frac{\varepsilon (s^* ,\rho ,0)E^2}{8\pi }+\ldots \ .
\label{eff17}
\end{eqnarray} 

For an adiabatic \begin{math} (ds^*=0) \end{math} incompressible 
\begin{math} (d\rho=0) \end{math} steady state Bernoulli flow, 
Eqs.(\ref{eff15}), (\ref{eff16}) and (\ref{eff17}) imply the central 
result of this section 
\begin{equation}
P+\frac{1}{2}\rho v^2 + \rho gz -\frac{\epsilon E^2}{8\pi}={\rm const}.
\label{eff18}
\end{equation}
In the case that \begin{math} {\bf E}=0  \end{math}, the Bernoulli 
Eq.(\ref{eff2}) indicates that water under the influence of gravity alone 
flows down a wall and may splash at the bottom. On the other hand, if the electric 
field on the bottom of the wall is negligible and the electric field on top of 
wall is large, then a polar dielectric fluid can crawl up a wall and may splash 
at the top. Such top splash processes on both of two beakers has been a precursor 
for building a water bridge. 

\section{Conclusions \label{conc}}

Water subject to high electric fields can sustain structures that are more 
than just a bit unusual. An electric field directed parallel  
to the water cylinder axis can create a tension as in a stretched rubber band but with 
different causes. In the case of the rubber band, the tension arises from the high 
entropy of random knotted polymer chains. In a polar liquid the tension arises 
out of long ordered chains of low entropy aligned coherent dipolar domains
\cite{DelGiudice:1988,DelGiudice:2006,Sivasubramanian:2005,Sivasubramanian:2003,Sivasubramanian:2002}. 
The resulting tension in the water bridge sustains a siphon tube between two beakers 
without the requirement that a new external siphon tube structure of other materials 
be introduced.

We shall conclude with some close analogies between a polar dielectric fluid, such 
as water, and a quantum superfluid such as liquid \begin{math} ^4_2He  \end{math}. 
In the liquid \begin{math} ^4_2He  \end{math} superfluid case, it is known that the 
superfluid film can act similarly to a siphon in which the film climbs over the wall 
and down the other side of the wall until the lowest possible gravitational energy 
is achieved. Similar flows can be induced in water by the application of high 
electric fields. In both cases, the flows are carried with little entropy 
production, although in the case of deionized water there is a mild Ohmic heating due 
to the transport of a few remaining ions. 

Finally, when the vorticity \begin{math} {\bf \Omega} \end{math} becomes 
important\cite{Chorin:1994}, 
as in the boundary layer between the wall and the Bernoulli flow, one may expect quantum 
vortex lines in water with the circulation condition\cite{Feynman:1955}   
\begin{math} \oint {\bf v}\cdot d{\bf r}=2\pi \hbar /m  \end{math}. 
The reasoning is that in a coherent liquid flow, the velocity potential 
\begin{math} \varphi \end{math} in Eq.(\ref{eff14}) determines the phase factor 
\begin{math} \exp [i(m/\hbar )\sum_j \varphi ({\bf r}_j,t)] \end{math} in the many 
body fluid wave functions. This quite general connection between the Bernoulli flow 
Eq.(\ref{eff14}) and the quantum mechanical phase is not self evident\cite{Widom:1990}. 
For this reason we have placed the mathematical details of the proof in the attached 
Appendix\ref{flow}.

\appendix
\section{\ Quantum Fluid Phase \label{flow}}

The velocity potential \begin{math} \varphi ({\bf r},t) \end{math} in the Bernoulli 
Eqs.(\ref{eff14}) and (\ref{eff17}), 
\begin{equation}
P+\rho \left[\frac{\partial \varphi}{\partial t}+
\frac{1}{2}|{\bf grad}\varphi |^2+\Phi +\varpi \right]=0,
\label{flow1}
\end{equation} 
may be employed in the phase of the many body fluid quantum wave functions,
\begin{eqnarray}
U(t)=\exp\left[\frac{im}{\hbar }\sum_j \varphi ({\bf r}_j,t) \right],
\nonumber \\ 
U(t)=\exp\left[\frac{i}{\hbar}\int \hat{\rho} ({\bf r})\varphi ({\bf r},t)d^3{\bf r}\right],
\label{flow2}
\end{eqnarray} 
wherein \begin{math} {\bf r}_j  \end{math} is the position of the 
\begin{math} j^{\rm th} \end{math} molecule and the quantum mechanical field operator for 
the mass density is given by\cite{Landau:1941} 
\begin{equation}
\hat{\rho} ({\bf r})=m\sum_j \delta ({\bf r}-{\bf r}_j).
\label{flow3}
\end{equation}
We presume a fluid microscopic Hamiltonian of the form 
\begin{equation}
\hat{H}=\frac{1}{2m}\sum_j |\hat{\bf p}_j|^2+V\ \ {\rm in\ which}
\ \ \hat{\bf p}_j=-i\hbar {\bf grad}_j
\label{flow4}
\end{equation}
and the positions \begin{math} \{{\bf r}_j\} \end{math} all commute with 
\begin{math} V \end{math}. 
The field operator for the mass current density is thereby\cite{Landau:1941} 
\begin{equation}
\hat{\bf J} ({\bf r})=\frac{1}{2}\sum_j \{\hat{\bf p}_j ,\delta ({\bf r}-{\bf r}_j)\}
\label{flow5}
\end{equation}
wherein the curly brackets indicate an anti-commutator; 
\begin{math} \{a,b\}\equiv ab +ba \end{math}. When viewed as a unitary 
transformation with the Bernoulli velocity potential \begin{math} \varphi ({\bf r},t) \end{math} 
determining the phase, one notes that 
\begin{eqnarray}
\hat{\rho }^\prime =U^\dagger \hat{\rho } U=\rho,
\nonumber \\ 
\hat{\bf J}^\prime =U^\dagger \hat{\bf J } U=\hat{\bf J }
+\hat{\rho }\ {\bf grad }\varphi ,
\label{flow6}
\end{eqnarray} 
the Bernoulli velocity \begin{math} {\bf v}={\bf grad }\varphi  \end{math} 
contributes to the current leaving the density unchanged. 

The unitary transformation on the Hamiltonian contains a time derivative term 
as given by 
\begin{eqnarray}
\hat{\cal H}=U^\dagger \hat{H} U-i\hbar U^\dagger \frac{\partial U}{\partial t}\ ,
\nonumber \\ 
\hat{\cal H}=\hat{H}+\int \hat{\bf J}\cdot {\bf grad}\varphi d^3 {\bf r}
\nonumber \\ 
+\int \hat{\rho}\left[\frac{\partial \varphi }{\partial t}
+\frac{1}{2}|{\bf grad}\varphi |^2\right]d^3 {\bf r},
\label{flow7}
\end{eqnarray}
leading finally to 
\begin{equation}
\hat{\cal H}= \hat{H}-{\cal E}+\int \hat{\bf J}\cdot {\bf grad}\varphi d^3 {\bf r}.
\label{flow8}
\end{equation}
In the final Hamiltonian Eq.(\ref{flow8}), the Bernoulli Eq.(\ref{eff14}) has been 
invoked with a total enthalpy  
\begin{equation}
{\cal E}=\int \rho (w+\Phi )  d^3 {\bf r}.
\label{flow9}
\end{equation}
The total enthalpy may be found in the microcanonical ensemble via the 
eigenvalue problem 
\begin{math} 
\hat{H}\left|n \right>={\cal E}_n\left|n \right> 
\end{math}. Diabatic entropy producing terms producing vorticity in boundary layers 
from the Bernoulli irroational flow arise from the interaction Hamiltonian  
\begin{equation}
\hat{H}_{\rm int}=\int \hat{\bf J}\cdot {\bf grad}\varphi d^3 {\bf r}.
\label{flow10}
\end{equation}

\end{document}